
%
\hsize=31pc
\vsize=55 truepc
\baselineskip=26 truept
\hfuzz=2pt
\vfuzz=4pt
\pretolerance=5000
\tolerance=5000
 \parskip=0pt plus 1pt
\parindent=16pt
\def\avd#1{\overline{#1}}
\def\avt#1{\langle#1\rangle}

\def\s{\sigma}
\def\P {{\cal{ P}}}
\def\Z {{\cal{ Z}}}
\def\s{{\sigma}}
\def\J {{\widetilde{ J}}}
\def\sp {{\widetilde{ \sigma}}}
\def\x {{\widetilde{ x}}}
\def\b {{\widetilde{ \beta}}}
\def\f {{\widetilde{ f}}}
\vskip 1.truecm\noindent
\centerline{\bf Mean field solution of the random Ising model}
\medskip
\centerline{\bf on the dual lattice}
\vskip 1.4truecm\noindent
\centerline{M. Serva$^{1,2}$, G. Paladin$^3$ and  J. Raboanary$^1$}
\vskip .4truecm
\centerline{$^{1}$ \it Institut Superieure Polytechnique de Madagascar}
\centerline{ \it Lot II H 31 Ter, Ankadindramamy
101 Antananarivo, Madagascar}
\vskip .7truecm
\centerline{\it $^{2}$Dipartimento di Matematica,  Universit\`a dell'Aquila}
\centerline{\it I-67100 Coppito, L'Aquila, Italy}
\vskip .7truecm
\centerline{\it $^{3}$Dipartimento di Fisica,  Universit\`a dell'Aquila}
\centerline{\it I-67100 Coppito, L'Aquila, Italy}
\vskip 1.6truecm
\centerline{ABSTRACT}
\vskip .4truecm
 We perform a duality transformation
 that allows one to express the partition function
 of the  $d$-dimensional Ising model with
random nearest neighbor  coupling
 in terms of new spin variables
defined on the square plaquettes of the lattice.
The dual model  is solved in the mean  field approximation.
\hfill\break
\hfill\break
\noindent
PACS NUMBERS: 05.50.+q, 02.50.+s
\vfill\eject
\vskip .4truecm

The Ising model with random coupling plays a fundamental role
 in the theory of disordered systems.
In this field, one of the major results
 is the Parisi solution of the infinite range model where at low temperature
 the system becomes a spin glass with a replica symmetry breaking [1].
 However, there is
 no exact solution for the model with nearest neighbor interaction,
 and it is still unclear whereas a glassy phase
 is present in three dimensions at finite temperature.

In this letter,
we perform a duality transformation
of the Ising model
 with random nearest neighbor coupling that assume
 the values $J_{ij}=\pm1$ with equal probability.
The model is thus defined on a dual lattice where the spin
variables are attached to the square plaquettes.
The advantage is that the non-linear part of the dual Hamiltonian
has constant coefficients instead of random ones.
It is therefore possible to use the standard mean field
approximation to estimate the quenched free energy.
However, in the dual lattice the ratio between number of spins
and number of links increases with the dimension
at difference with what happens in the original lattice
so that the mean field approximation
 becomes worst at increasing the dimensionality.
Our solution is thus optimal in two dimensions
 where it gives a rather good estimate of the ground state energy.

The partition function
 of the $d$-dimensional Ising models
 on a  lattice of $N$ sites
 with nearest neighbour couplings $J_{ij}$ which are
independent identically distributed random variables,
 in absence of external magnetic field, is
$$
Z_N(\beta, \{J_{ij}\})  =  \sum_{ \{ \sigma \} }  \prod_{(i,j)}
\exp (\beta J_{ij} \sigma_i \sigma_j )
\eqno(1)
$$
where the sum runs over the spin configurations $\{\s\}$, and
 the pruduct over  the nearest neighbor sites $(i,j)$.
One is interested in computing the quenched free energy
$$
f=- \lim_{N \to \infty} {1\over \beta N} \avd{\ln Z}
\eqno(2)
$$
where $\avd{A}$ indicates the average of an observable $A$
 over the distribution of the random coupling.
The quenched free energy is a self-averaging quantity,
 i.e. it is obtained in the thermodynamic limit
 for almost all realizations of disorder [1].
Even in one dimension, it is difficult to find an exact solution
for $f$ in presence of an external constant magnetic field [2].
  On the other hand, it is trivial to compute
the so-called annealed free energy
$$
f_a=-\lim_{N \to \infty} {1\over \beta N} \ln \avd{ Z} \ ,
\eqno(3)
$$
corresponding to the free energy of a system where the random coupling
 are not quenched but can thermalize with a relaxation time comparable
 to that one of the spin variables.
An easy calculation shows that in our case
$$
f_a= -{1\over \beta} \, (\, \ln 2 +d \ln \cosh \beta \, )
\eqno(4)
$$
However, $f_a$ is a very poor approximation of the quenched free energy,
 and is not able to capture the qualitative features of the model.

In order to estimate (1),
 it is convenient to use the link variable $x_{ij}=\s_i \, \s_j$,
since only terms corresponding to products of the variables $x_{ij}$
 on close loops survive after summing over the spin configurations:
 on every close loop of the lattice $\prod x_{ij}=1$,  while
 $\prod x_{ij}=\s_a \, \s_b$ for a path from the site $a$ to the site $b$.
   A moment of reflection shows that it is sufficient to fix
 $\prod x_{ij}=1$ on the
 elementary square plaquettes $\P$
 to automatically fix  it on all the close loops.
The partition function thus becomes
$$
Z_N(\beta, \{J_{ij}\})  =  \sum_{ \{ x_{ij} \} }   \
\prod_{i=1}^{N_p}  {1+\x_i\over 2} \, \prod_{(i,j)}  e^{\beta J_{ij} x_{ij}}
\eqno(5)
$$
where the number of plaquettes is $N_p=d(d-1)N/2$, and
 we have introduced the plaquette variable
$\x_i=\prod_{\P_i} x_{ij}$.

For dichotomic random coupling $J_{ij}=\pm 1$ with equal probability,
 the free energy of the model is invariant under the gauge transformation
$x_{ij} \to J_{ij} \, x_{ij}$, so that one has
$$
Z_N  =  \sum_{ \{ x_{ij} \} }   \
\prod_{i=1}^{N_p}  {1+\x_i \J_i \over 2} \, \prod_{(i,j)}  e^{\beta x_{ij}}
\eqno(6)
$$
where $\J_i=\prod_{\P_i} J_{ij}$ is again a dichotomic random variable
 (the `frustration' [3] of the plaquette $\P_i$).
It is worth remarking that (6) gives the partition function in terms of a sum
 over the $2^{dN}$  configurations of the independent random variables
 $\x_{ij}=\pm 1 $   with probability
$$
p_{ij}= {e^{\beta x_{ij}} \over 2  cosh  \beta}
\eqno(7)
$$
 In the following we shall indicate
the average of an observable $A$
 over such a normalized weight  by $\avt{A}$, e.g
 $\avt{x_{ij} }=\tanh \beta$ and $\avt{\x_i}=\tanh^4 \beta$.
With such a notation, the partition function assumes the compact form
$$
Z_N= 2^{(dN-N_p)} \, \cosh^{d N} (\beta)
  \ \avt{ \, \prod_{i=1}^{N_p}(1 + \x_i \J_i) \, }
\eqno(8)
$$
Now comes the key step. We estimate the average in (8) by a geometrical
construction.
Let us introduce the dual lattice [4]
 as the lattice whose sites are located at the centers of each square
  of the original lattice.
A dual spin variable is attached to each
 square plaquette and can assume only the values $\sp_i=\pm 1$
 with equal probability,
 so that one has the identity
$$
(1 + \x_i \J_i) = \sum_{\sp_i=\pm1} (\x_{i}\, \J_i)^{(1+\sp_i)/2}.
\eqno(9)
$$
Since there is a one-to-one correspondence between links on the original and
on the dual lattice, we can compute the link-average
noting that
$$
\avt{ \prod_{i=1}^{N_p} \x_i^{{(1+\sp_i)\over 2} }} =
   \avt{ \, \prod_{(i,j)}
x_{ij}^{ \sum_{k \in (i,j)} { (1+\sp_k)\over 2 } }
  } =
\prod_{(i,j)} \tanh^{1-P_d^{(i,j)}}(\beta)
\eqno(10)
$$
where
$\sum_{k \in {(i,j)} }$ is the sum on the
 $2(d-1)$ nearest neighbor plaquettes variables
 with a common link $(i,j)$ and for sake of simplifying the notation
we have defined the new link variable
$$
P_d^{(i,j)}=\prod_{k \in {(i,j)}} \sp_k
\eqno(11)
$$
Inserting (10) and (9) into (8) one has
$$
Z_N=2^{(dN-N_p)} \, \cosh^{d N} (\beta)
 \sum_{\{ \sp \}} \prod_{i=1}^{N_p} \J_i^{(1+\sp_i)/2}
\, \prod_{(i,j)} e^{\b (P_d^{(i,j)} -1) }
\eqno(12)
$$
where
the inverse temperature of the dual model is
 $$
\b= - \, {1\over 2} \, \ln \, \tanh (\beta)
\eqno(13)
$$
that  vanishes when the temperature $T=\beta^{-1} \to 0$.
The quenched free energy (2) thus is
$$
-\beta \, f(\beta)={d\over 2} \, \left(
   (2-d)  \ln 2 + \ln \sinh(2 \beta)
 -  (d-1)  \, \b \, \f(\b) \, \right)
\eqno(14)
$$
where the free energy of the dual model is
$$
\f(\b)= - \lim_{N_p \to \infty}
 {1\over\b  N_p} \ln \Z_{N_p}
\eqno (15)
$$
with
$$
\Z_{N_p}=\sum_{\{\sp_i\}} \exp  \left(\b \sum_{ (i,j)}
 P_d^{(i,j)}\right) \ (-1)^{K(\{\sp_i\},\{\J_i\})}
\eqno(16)
 $$
and
$$
K=\sum_{i=1}^{N_P} \left(
 {1+\sp_i\over 2\pi \, i}\right)  \, \ln (\J_i)
\eqno(17)
$$
Since  the Hamiltonian of the dual model
is defined by the relation $\Z_{N_p}=\sum_{\{\sp\}}  e^{-\b H}$,
from (16) one sees that
$$
H=
  - \sum_{ (i,j)} P_d^{(i,j)} \
  - \,  \sum_{i=1}^{N_p} \, \ln(\J_i) \,  {(1+\sp_i)\over 2 \b}
\eqno(18)
$$
Let us stress that the non-linear term of the dual Hamiltonian (18)
does  not depend on the random coupling like in the original model and
 the randomness enters via the presence of a sign. In fact,
 the weight $\exp(-\b \, H)$  does not defines
 a standard Gibbs probability measure on the dual lattice:
  it defines  a signed probability measure.
  For instance, in $2d$,  $P_{d}^{(i,j)}=\sp_i \, \sp_j$
 so that the Gibbs measure of a configuration of
 the dual random Ising model
 differs from that one of the pure Ising model only
 for the presence of the random sign  determined by the index $K$
related to the frustrations of the square plaquettes $\{\J_i\}$.

This is the first result of our letter. Its
  importance stems from the fact that
 it is now possible to linearize the Hamiltonian (18) by
 introducing the magnetization
$$
m=\lim_{N_p \to \infty} {1\over N_p} \sum_i^{N_p} \sp_i
\eqno(19)
$$
Indeed, if we neglect the fluctuations,  the non-linear
  term of (18) can be estimated as
$$
\sum_{(i,j)} P_d^{(i,j)}=d \, N \, m^{2(d-1)}
$$
 so that  (16) becomes
$$
\Z_{N_P}=\sum_{\{\sp\}}\ \prod_{i=1}^{N_p} \J_i^{(1+\sp_i)/2}
 \exp( \, \b \,  d \, N \, m^{2(d-1)})
\eqno(20)
$$
Let us find the mean field solution, by using an auxiliary field
 $\Phi$. Recalling the saddle point method one immediately sees
 that in the limit $N \to \infty$,
$$
e^{N \, d \, \b  \, m^{2(d-1)} }\sim \int_{-\infty}^{\infty}
  d\Phi \exp \left(
 N \b d \,
 ( C_d \, m \Phi^{2d-3} - \Phi^{2(d-1)} )\right)
\eqno(21)
$$
where the constant $C_d$  is determined by maximizing the
 argument of the exponential and reads
$$
C_d= 2 \,(d-1) \, ( \, 2d-3 \, )^{-(d-3/2)/(d-1)}
\eqno(22)
$$
As a consequence one can write the partition function as
$$
\Z_{N_P}\sim \max_{\Phi}   \ \exp \left(-N \b d \Phi^{2(d-1)} \, \right) \,
 \sum_{\{\sp\}}\prod_{i=1}^{N_p} \J_i^{(1+\sp_i)/2}
 e^{\b d C_d \Phi^{2d-3}\sp_i}
\eqno(23)
$$
Now the effective Hamiltonian is linear
 and one can explicitly carry out the
 sum over the $N_p$ independent random variables $\sp_i$,
 obtaining
$$
-\b \f(\b)= d \,
 \max_{\phi} \left( \,
 {(d-1) \over 4} \ln(2 \sinh(\gamma_d \, \b \Phi^{2d-3}) )
 - \b \, \Phi^{2(d-1)} \right)
\eqno(24)
$$
with $\gamma_d = 2   \, C_d  \, d $. The maximum is realized
 by the value of $\Phi^*$ that is the solution of the self-consistency equation
$$
{(2d-3) \gamma_d \over 8} \ \coth(\gamma_d \b \Phi^{2d-3})= \Phi
\eqno(25)
$$
In $2d$, (25) assumes the simple form $ \coth(8 \b \Phi)=\Phi$.
The graphical solution of this implicit equation is showed
 in fig 1. One sees that $\Phi^*$ should be
always larger than unity
and at $\b \to \infty$ (infinite temperature $T=\beta^{-1}$ limit)
$\Phi^*=1$. It can appear rather odd that
 in the dual model  the magnetization $\Phi^* \ge 1$.
 This stems from the fact that
 the Gibbs probability measure $exp(-\b H)$ is
 a signed measure
 because the random coupling is transformed into
 a complex random magnetic field in (15).
 From fig 1, it is also clear that the
 mean field solution does not exhibit  phase transitions
at finite temperature. However there is an
 essential singularity at $T=0$, since inserting (24) into (15) and (14)
 one sees that $f \sim \exp(1/T)$ for $T \to 0$.

It is important to stress that the mean field solution
  does not improve at increasing the dimensionality,
 since the ratio $N_p/N$ between number of plaquette spins
and number of links in the dual lattice  increases as $(d-1)$
at difference with the standard Ising model
 where the ratio of spins over links decreases as $d^{-1}$.

It is possible to explicitly
 solve the self-consistency equation for $\b \to 0$
where (25) becomes
$$
\Phi^* \sim  ( \, 12 \, \b \, )^{-1/2}
\qquad {\rm  for} \  d=2
\eqno(26a)
 $$
and
$$
\Phi^* \sim \left(2d-3 \over 8\right)^{1/(2d-2)} \, \b^{-1/(2d-2)}
\qquad {\rm  for} \  d \ge 3
\eqno(26b)
 $$
Such a relation shows that when $d \to \infty$
 one has $\Phi^* \to 1$  for $T \to 0$
and then for all $T$'s.
 The high dimension limit  is therefore trivial.
The mean field approximation works at its best in two dimensions.
For instance, the zero temperature energy of the mean field solution
is $E_0=-3d/4$ which is a fair estimate in $2d$
where the numerical simulations [5] give $E_0=-1.404 \pm 0.002$.
In fig 2, we show the free energy as a function
 of $T$ in $2d$ compared with the annealed free energy (4).
 One sees that entropy is negative at low temperature,
  thus indicating that the solution is unphysical.
 As a consequence, a better estimate of the ground state energy
 is given   by the maximum of $f(\beta)$, following a standard argument of
 Toulouse and Vannimenus [6],
and one has $E_0 \le \max_{\beta} f(\beta)=-1.468$.

In conclusion, we have obtained two main results.
 (1) Formulation of the random coupling Ising model on the dual
 lattice made of square plaquettes. The dual model has signed
 Gibbs probability measure and magnetization larger than unity.
 (2) Application of the mean field approximation
 to solve the dual model. The approximation is sensible
 at low dimension.

In our opinion, the mean field
 approach is very promising at least in two dimensions.
It has a good heuristic power and  there
 are still a lot of open problems in its framework,
 such as finding a Ginzburg-Landau criterion
 or refining the mean field approximation in  a cluster expansion scheme.
\bigskip
\bigskip
We acknowledge the financial support  of the I.N.F.N.,
  National Laboratories  of Gran Sasso.
\vfill\eject

\vfill\eject
\noindent
{\bf References}
\bigskip
\bigskip
\bigskip
\item{[1]}
M. Mezard, G. Parisi and M. Virasoro, {\it Spin glass theory and beyond},
 World Scientific Singapore 1988
\bigskip
\item{ [2]}  G. Paladin and M. Serva, Phys. Rev. Lett. {\bf 69}, 706 (1992).
\bigskip
\item{ [3]}
G. Toulouse, Commun. Phys.  {\bf 2}, 115 (1977)
\bigskip
\item{ [4]}
H. A. Kramers and G. H. Wannier,
Phys. Rev. {\bf 60}, 252 (1941)
\bigskip
\item{[5]}
L. Saul and M. Kardar,
Phys. Rev. E {\bf 48} (1993) 48
\bigskip
\item{ [6]}
G. Toulouse and J. Vannimenus, Phys. Rep. {\bf 67}, 47 (1980)

\vfill\eject
\vskip 0.8truecm
\centerline {\bf Figure Captions}
\vskip 0.5truecm
\noindent

\item{Fig. 1}
Graphical solution of the implicit equation (23) in $2d$,
 at $T=\beta^{-1}=1$  corresponding to  $\b=0.136..$.
 The full lines are $\coth(8 \b \Phi)$ versus $\Phi$
 and the straight line $\Phi=\Phi$.

\bigskip
\item{Fig. 2}
Random Ising model in $2d$:
the annealed free energy $f_a$ (dashed line) and the
mean field solution (full line)  versus temperature $T=\beta^{-1}$.
The dashed lines are the Maxwell constructions obtained by imposing
that the free energy is a monotonous non-decreasing function of $T$.
 The annealed solution gives a ground state energy
$E_0=-1.559$; the mean field solution gives
 $E_0=-1.468$; the numerical result of [5] is $E_0=-1.404\pm 0.002$
\bye